\documentclass[a4paper,fleqn,usenatbib,useAMS]{mnras}
\usepackage{epsfig,amssymb,amsmath,rotating,lineno,graphicx,hyperref}
\usepackage{newtxtext,newtxmath}
\usepackage{booktabs,chemformula}

\usepackage{amssymb}	
\usepackage[utf8]{inputenc}
\usepackage{graphicx}
\usepackage{xcolor}
\usepackage{xurl}
\usepackage{lipsum}
\usepackage{booktabs}
\usepackage{amsmath}
\usepackage{seqsplit}
\usepackage{float}
\usepackage{array}
\usepackage{subcaption}
\usepackage[export]{adjustbox}
\usepackage{auto-pst-pdf} 
\usepackage{wrapfig}
\usepackage{hyperref}
\usepackage{natbib}
\usepackage[para]{threeparttable}
\usepackage{caption}
\usepackage{multirow}
\usepackage{caption,stackengine}
\usepackage[T1]{fontenc}
\usepackage{ae,aecompl,arydshln}
\usepackage{mathtools}
\newcommand\asat{{\it AstroSat}}

\title[Origin of EFB of Z sources]{Probing the origin of the extended flaring branch of Z-type X-ray binaries GX~340+0 and GX~5-1 using \em{AstroSat}}

\author[Dutta et al.]{Tanmoy Dutta$^{1}$, Mayukh Pahari$^{1}$, Anish Sarkar$^{1}$, Sudip Bhattacharyya$^{2}$, Yash Bhargava$^{2}$ \\
$^{1}$ Department of Physics, Indian Institute of Technology, Hyderabad, Kandi, Sangareddy 502285, India \\
$^{2}$ Department of Astronomy and Astrophysics, Tata Institute of Fundamental Research, 1 Homi Bhabha Road, Colaba,
Mumbai 400005, India}

\begin{document}

\pagerange{\pageref{firstpage}--\pageref{lastpage}} \pubyear{2024}

\maketitle

\label{firstpage}

\begin{abstract}
`Z' type neutron star low-mass X-ray binaries typically show a `Z'-like three-branched track in their hardness intensity diagram.
However, a few such `Z' sources show an additional branch known as the extended flaring branch (EFB). EFB has been poorly studied, and its origin is not known. It is thought to be an extension of the flaring branch (FB) or associated with Fe K$\alpha$ complex or an additional continuum due to the radiative recombination continuum (RRC) process. 
Using \asat{} observations, we have detected the EFB from two `Z' sources, GX 340+0 and GX 5--1, and performed a broadband spectral analysis in the 0.5--22 keV energy range. During EFB, both sources show the presence of a significant RRC component with absorption edges at $7.91^{+0.16}_{-0.15}$ keV and $8.10^{+0.16}_{-0.17}$ keV, respectively along with blackbody radiation and thermal Comptonisation. 
No signature of RRC was detected during the FB, which is adjoint to the EFB. No Fe K$\alpha$ complex is detected. Interestingly, inside EFB dips of GX 5-1, for the first time, we have detected flaring events of 30--60s, which can be modelled with a single blackbody radiation. 
During the FB to EFB transition, an increase in the blackbody radius by a factor of 1.5--2 is observed in both sources. Our analysis strongly suggests that EFB is not an extension of FB or caused by the Fe K$\alpha$ complex. Rather, it is caused by a sudden expansion of the hot, thermalised boundary layer and subsequent rapid cooling.
\end{abstract}


%%%%%%%%%%%%%%%%% BODY OF PAPER %%%%%%%%%%%%%%%%%%

\section{INTRODUCTION}
\label{sec:1}
Based on the behaviour in the hardness intensity diagram (HID), an accreting Neutron star X-ray binary (NSXB) source can be classified into two classes: `Z'-type and atoll source \citep{hasinger}. Such a distinction was primarily attributed to the different mass accretion rates and spectroscopic evolution in the two different systems. There are mainly three spectral branches that we see in the HID of the Z-type sources: horizontal branch (HB), normal~branch (NB), and flaring branch (FB)\citep{hasinger}. The sources are observed to spend an arbitrary duration of time at a particular branch but show a continuity while evolving along the Z-track \citep[without jumping across different branches;][]{yash2023,pahari2024astrosat}.  The transition between the different branches, especially from HB to NB, is widely believed to be due to the radio jets and outflows \citep{miglairi}. There is no clear consensus on which physical model describes the spectral evolution of different branches. But spectral evolution is usually described as a combination of soft thermal emission from the accretion disc, a black body emitting component known as the boundary layer and a hard, non-thermal Compton component \citep{homan,yash2023}. 

In `Z' sources, very seldom, a fourth branch, known as the extended flaring branch (EFB) \citep{church,gibiec}, is observed at the end of the FB. Despite a few efforts, the origin of such a branch is still unclear. \citet{hasinger90} conducted a comprehensive study of Cyg X-2 across multiple wavelengths using {\it Ginga}. They observed intensity decreases on the FB, which were interpreted as absorption dips. \citet{kuulker95} proposed that in Cyg X-2 and GX~340+0, the FB in colour-colour representations correlates with X-ray dipping, contrasting to the behaviour observed in Sco-like sources where the FB corresponds to significant increases in intensity during flaring episodes. They suggested that Cyg-like sources have higher inclination angles compared to Sco-like sources, proposing a model to explain the differences between the two involving absorption or scattering within the inner disc. However, reliable spectral fitting of the EFB was lacking in establishing its nature definitively.
Using {\it RXTE} observations, EFB has been reported from GX~340+0 \citep{pennix} and GX~5-1 \citep{jonker2000power}. Both sources have been chosen here to study the origin of EFB using \asat{} broadband spectral analysis.      
GX~340+0, a luminous, low mass neutron star X-ray binary, traces a `Z' shaped track in HID \citep{hasinger}. Previously, \citet{yash2023} have thoroughly performed HB and NB branch-resolved spectral analysis of GX 340+0 using \asat{} in 0.8--25 keV. They have found that HB, NB and hard apex can be modelled by the emission from the neutron star surface, accretion disk and compromising corona covering the inner disk/boundary layer. With the launch of the Imaging X-ray Polarimetry Explorer (IXPE), there has been significant improvement achieved in understanding X-ray emission geometry of Z-type NSXBs using spectro-polarization measurements \citep{cocchi23,soffitta2024,yash2024,Monaca2024Highly,fabiani24,yu24}. Among them, \citet{fabiani24} discovered variable polarization in GX~5-1: 3.7$\pm$0.4\% in HB while 1.8$\pm$0.4\% during NB/FB using a multi-wavelength campaign. Similarly, \citet{yash2024} has discovered significant X-ray polarization (4.02 $\pm$ 0.35\%)in the HB of GX~340+0 while a weaker polarization of 1.22 $\pm$ 0.25\% is detected during NB in GX~340+0 \citep{yash24nb}.

Discovered in 1968, GX 5-1 is the second brightest LMXB (after Sco X-1) \citep{fisher1968}. Using {\it Ginga} observations, the source is classified as the `Z' source by \citet{hasinger}. In the analysis of {\it Exosat} and {\it Ginga} data, the flaring branch was discovered for the first time by \citet{kul1994}. The first EFB was observed in the source by \cite{church} using the data from {\it RXTE} satellite. The branch-resolved timing analysis of GX 5-1 was carried out using \asat{} observations \citep{Bhulla_2019}. 
Using \textit{RXTE} data \citet{church} found that the EFB is an extension of the flaring branch, which maintains similar characteristics as per EFB. The mass accretion rate continues to fall, and unstable nuclear burning continues. During EFB, \citet{church} detected strong emission lines at energies between 7.8--9.4 keV, suggesting radiative recombination continuum (RRC) of Fe XXVI at 9.28 keV and of lower energy states.
During RRC, the photon emitted during the recombination of an electron and an ion has the same energy as the kinetic energy of the electron. Therefore, the emitted spectrum is a continuum in nature with a sharp edge at the binding energy level \citep{tucker1966}. The detection of RRC in X-ray spectra has been reported few times in other sources, e.g., using {\it Suzaku} observations, \citet{ozawa2009} detected a line-like excess around Fe Ly$\alpha$ and saw-edge-shaped bump around 8 keV from the Galactic supernova remnant (SNR) W49B while \citet{yamaguchi2009} observed the same from another SNR IC 443. The detection from both SNRs is highly suggestive of an overionized plasma with a large fraction of H-like ions. By analyzing \textit{XMM-Newton}/EPIC spectra using \texttt{redge} model \citet{sugawara2008} detected RRC structure around 0.49 keV from the Wolf–Rayet binary $\theta$ Muscae. Using X-ray spectral analysis of SNR ejecta \citet{greco2020} observed that the RRC component appeared when the plasma was made of pure metal ejecta. Although X-ray spectral features observed during EFB of Z-type NSXBs are similar to what is predicted by RRC, the origin is unclear.
Investigating the nature of EFB in Cygnus X~2, \cite{gibiec} found that the extended flaring branch is not the continuation of the flaring branch; rather, it is due to the absorption of the flaring branch due to the outer layer of the accretion disk. Therefore, the origin of EFB remains unclear.
To shed further light on such an unsettled issue, we have thoroughly analyzed the extended flaring branch of the two Cyg-like sources, GX~5-1 and GX~340+0, using the simultaneous \textit{LAXPC} and \textit{SXT} observations on-board \asat{} using broadband energy range of 0.3--30.0 keV. We have separately analysed spectra exclusive to FB and EFB and have compared their properties. Interestingly, we have detected flaring events within EFB, which can be described by a single blackbody emission with a blackbody emission significantly higher than that observed during FB. We observed that the EFB is not a continuation of the flaring branch but rather due to the new spectral component known as radiative recombination continuum (RRC). Such a component is different than Fe K$\alpha$ complex and absent in FB.

The observation and data reduction for \textit{LAXPC} and \textit{SXT} are provided in Section \ref{sec:2.1} and \ref{sec:2.2}, respectively, while the analysis procedures for GX 5-1 and GX 340+0 are detailed in Sections \ref{sec:3.1} and \ref{sec:3.2}, respectively. Results of our analysis during FB, EFB and flaring within EFB for GX 5-1 are discussed in Section \ref{sec:4.1} while the FB and EFB results for GX 340+0 are provided in Section \ref{sec:4.2}. Detection significance testing of the new spectral component over continuum is presented in Section \ref{sec:5}. We have discussed the implications of our results in Section \ref{sec:6}.

\section{Observation and Data Reduction}
\label{sec:2}
The data has been obtained through India’s broadband X-ray observatory, \asat{} \citep{singh-2014}, using simultaneous observations from both large area X-ray proportional counter (\textit{LAXPC}) and soft X-ray telescope (\textit{SXT}) instruments. \textit{LAXPC} and \textit{SXT} observation details of GX~340+0 and GX~5-1 are provided in Table~\ref{tab:1}.

\subsection{LAXPC} 
\label{sec:2.1}
\textit{LAXPC} \citep{l-yadav, l-antia} consists of 3 independent but identical detectors, giving a collecting area of $\sim$6000 $cm^2$ at 15 keV. Its operational energy range is 3--80 keV. 
The \textit{LAXPC} data has been reduced using \textsc{LaxpcSoft v21June2023} with suitable response files for corresponding \textit{LAXPC} units. Cleaned event files and good time intervals (GTIs) for each satellite orbit have been created using \textit{LAXPC10} and \textit{LAXPC20} units. Due to poor data quality and calibration issues, data from \textit{LAXPC}30 is excluded from all further analyses. GTIs have been cleaned further to remove segments with poor data quality caused by telemetry losses or other factors like the initial and the final 100s transition in and out to the South Atlantic Anomaly (SAA) region, respectively. Such a screening accounts for nearly 5\% of total effective exposure. Lightcurves in 6--10 keV, 10--20 keV and 6--20 keV energy bands are extracted using the cleaned GTIs, and hardness intensity diagrams are computed for both sources GX~5-1 and GX~340+0.

\subsection{SXT}
\label{sec:2.2}
\textit{SXT} onboard \asat{} is a focusing X-ray telescope utilizing a charge-coupled device capable of X-ray imaging in the energy range of 0.3--7.0 keV with medium resolution \citep{singh-2017}. For both sources, we have used data from orbits similar to \textit{LAXPC}. \textit{SXT} data has been reduced using \textsc{SXTPIPELINE v1.5b}\footnote{\url{http://www.tifr.res.in/~astrosat_sxt/sxtpipeline.html}}, and \textsc{xselect v2.4g}. We have used an annular region in the image for spectral extraction to avoid the pileup issue. The annular region consists of two concentric circles with radii of 5 arcmin and 15 arcmin, respectively. Using \textit{LAXPC} GTI files, which correspond to FB and EFB, spectra and light curves have been extracted for both sources during FB and EFB. Orbit-specific \texttt{arf} files are generated using \textsc{sxt\_ARFModule\_{v02}} tool for spectral fitting.

\begin{table*}
\centering
\caption{Flaring and Extended flaring branch observation details of GX~340+0 and GX~5-1 with \asat{}.}

  \begin{tabular}{ccccccccc}
    \toprule
     & Source  & Branch  & Instrument & Observation& Orbit & Observation & Start & Exposure\\
     & name & name &  & ID & number & date & time & time\\
  \break &  &  &  &  &  &  & (HH:MM:SS)  & (s)\\
   \midrule 
        \label{val:18} & GX~5-1 & FB  & \textit{SXT} & T01\_056T01\_9000000356 & 2343 & 31-07-2017 & 07:48:01& 239 \\
    \label{val:18} &  & FB  & \textit{LAXPC} & T01\_056T01\_9000000356 & 2343 & 31-07-2017 &07:48:01& 5542 \\
    \label{val:18} &  & EFB  & \textit{SXT} & T01\_056T01\_9000000356 & 2343 & 31-07-2017 & 07:48:01& 111 \\
    \label{val:18} &  & EFB  & \textit{LAXPC} & T01\_056T01\_9000000356 & 2343 & 31-07-2017 &07:48:01 &  715 \\
    \midrule
    \label{val:18} & GX~340+0 & FB  & \textit{SXT} & G07\_016T01\_9000001420 & 9953 &04-03-2016 & 13:40:38& 2933\\
    \label{val:14} &  & FB & \textit{LAXPC} & G07\_016T01\_9000001420 & 9953 &04-03-2016 &13:40:38 & 4744 \\
    \label{val:14} &  & EFB & \textit{SXT} & G07\_016T01\_9000001420 & 9953 & 04-03-2016 &13:40:38 & 456 \\
    \label{val:14} &  & EFB & \textit{LAXPC} & G07\_016T01\_9000001420 & 9953 & 04-03-2016 &13:40:38 & 657.5 \\

    \bottomrule
  \end{tabular}
  \label{tab:1}
\end{table*}

\section{Data analysis}
\subsection{GX~5-1}
\label{sec:3.1}
\begin{figure*}
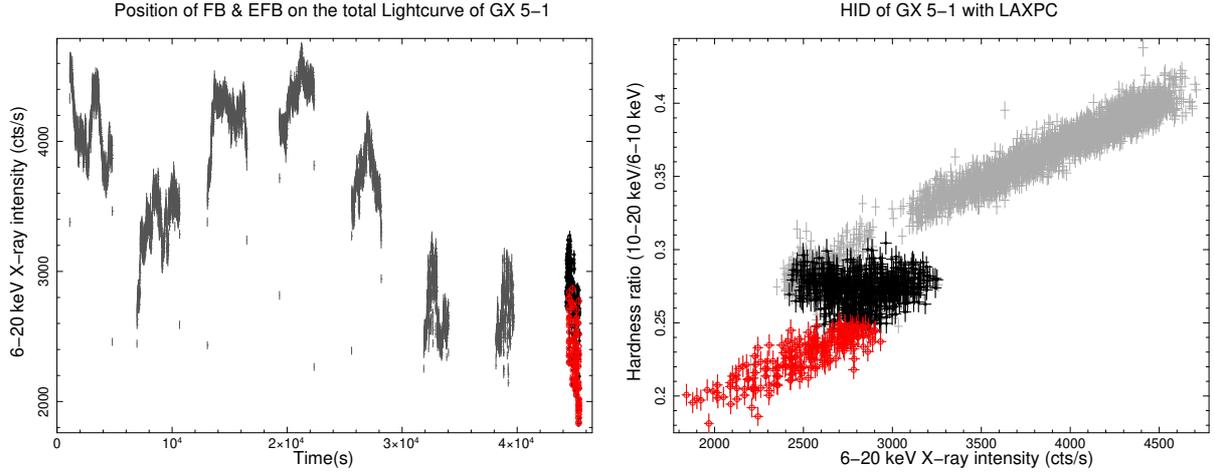

    \centering
\includegraphics[angle=-90,scale=0.33]{fig1a.eps}
\includegraphics[angle=-90,scale=0.33]{fig1b.eps}
    \caption{Left: \asat{}/LAXPC light curve of GX~5-1 in 6-–20 keV with 5s binsize. Right: the hardness–intensity diagram (HID) of the source with hard colour computed using the background-subtracted count rates in 10–-20 keV divided by that in 6–-10 keV and intensity represented by the total count rate in 6–-20 keV. The Z track has been divided into three zones with roughly similar extents on the Z track. Different colours have been used for points corresponding to different zones. The light curve of the left panel is coded with the same colours, which shows how the source moved from one zone to another on the Z track. See Section \ref{sec:4.1}.}
    \label{gx51-light-hid}
\end{figure*}

To identify the FB and EFB, we have used data from all orbits (2318-2344) for the observation ID T01\_056T01\_9000000356. Using the standard \textit{LAXPC} analysis procedure, we have extracted the background-subtracted light curve in 6--20 keV, combining \textit{LAXPC10} and \textit{LAXPC20}. HID is computed using all orbit data where hardness is defined as the ratio of the count rate in the energy range 10--20 keV to 6--10 keV, and intensity is defined as the count rate in 6--20 keV. HID and lightcurves are shown in Figure~\ref{gx51-light-hid}. Different branches like NB, FB and EFB are visible. Since the current study focuses on FB and EFB, they are marked by pluses and circles in both HID and light curves. The EFB branch is similar and parallel to the NB, but it has a lower hardness. 

For further studies, we have extracted the spectra of FB and EFB from GX~5-1 using \textit{LAXPC20} for the energy range 3--22 keV. Due to instrument gain instability and poor channel-to-energy calibration, \textit{LAXPC10} spectrum is excluded from further analysis. To match the energy resolution of \textit{LAXPC20} (approximately 15\%), we used optimal settings to have three energy bins per resolution or at a 5\% level.  Background spectra are extracted using the GTI, the same as the source spectrum. Suitable response files are used for spectral fitting. \textit{SXT} spectra simultaneous to \textit{LAXPC20} were extracted in the energy range 0.3--7 keV. \textit{SXT} spectra are binned such that each bin has a minimum of 30 counts. \textit{SXT} and \textit{LAXPC20} spectra of FB and EFB are jointly fitted with a suitable model using \texttt{XSpec version 12.14.0h} \citep{arnaud1996}.

\subsection{GX~340+0}
\label{sec:3.2}
To identify different branches from GX~340+0, we have used data on 04 March 2016 from all \asat{} orbits (9939-9959) of the observation ID \textit{G07\_016T01\_9000001420}.
We have extracted the spectrum of the FB and EFB using appropriate GTI for the \textit{LAXPC20} unit in the energy range of 3--22 keV. Similar to GX~5-1 analysis, spectra are grouped into three energy bins per resolution. Similar to GX~5-1, \textit{SXT} spectra are extracted and fitted jointly with \textit{LAXPC20} using similar models in \texttt{XSpec v 12.14.0h} \citep{arnaud1996}.

\begin{figure*}
    \centering
\includegraphics[angle=-90,scale=0.33]{fig2a.eps}
\includegraphics[angle=-90,scale=0.33]{fig2b.eps}
\includegraphics[angle=-90,scale=0.33]{fig2c.eps}
\includegraphics[angle=-90,scale=0.33]{fig2d.ps}
    \caption{Upper Left: \asat{}/LAXPC light curve of GX~5-1 of FB \& EFB in 6–-20 keV with 5 s bins. Upper Right: the \textit{LAXPC} hardness–intensity diagram (HID) of the GX~5-1 of FB \& EFB  with hard colour computed using the background-subtracted count rates in 10-–20 keV divided by that in 6-–10 keV and intensity represented by the total count rate in 6-–20 keV. Bottom left: Variation of the Hardness with respect to time. Bottom right: The portion of 0.3--8 keV \asat{}/SXT lightcurve, which is simultaneous with LAXPC, is shown. See Section \ref{sec:4.1}.}
    \label{fb-efb-5-1}
\end{figure*}

\begin{figure*}
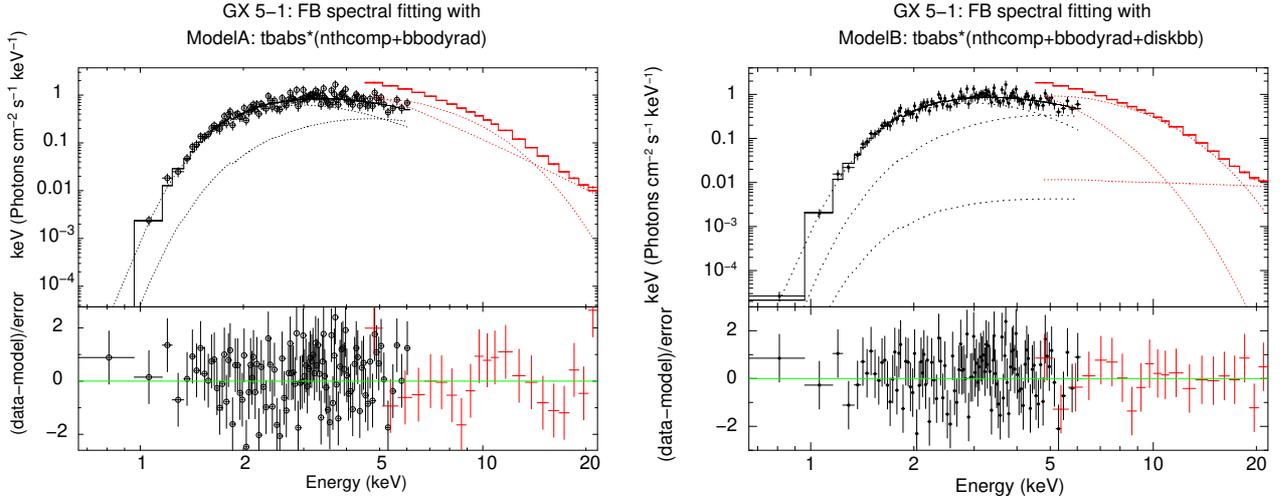

    \centering
\includegraphics[angle=-90,scale=0.32]{fig3a.eps} 
\includegraphics[angle=-90,scale=0.32]{fig3b.ps}
\caption{Best-fitting spectra of the flaring branch of GX~5--1 along with residuals. Left: The best-fittings model is ModelA: \texttt{tbabs}*(\texttt{nthcomp} + \texttt{bbodyrad}). Right: The best-fitting model is ModelB: \texttt{tbabs}*(\texttt{nthcomp} + \texttt{bbodyrad} + \texttt{diskbb}). The lower panel of both figures represents the residual of the best-fitting model. Dotted lines in each panel show individual model components. See Section \ref{sec:4.1.1}.}
    \label{fig:5-1-fb}
\end{figure*}

\begin{figure*}
    \centering
\includegraphics[angle=-90,scale=0.31]{fig4a.ps}  
\includegraphics[angle=-90,scale=0.31]{fig4b.ps}
    \caption{Best-fitting spectra of the extended flaring branch of GX~5-1 along with residuals: Left: To compare with the best-fitting model for FB spectra, ModelB: \texttt{tbabs}*(\texttt{nthcomp} + \texttt{bbodyrad} + \texttt{diskbb}) is used. A strong residual is observed near 9--10 keV. Right: best-fitting spectra and residual when the same spectra from the EFB of GX~5-1 are fitted with ModelC: \texttt{tbabs}*(\texttt{nthcomp} + \texttt{bbodyrad} + \texttt{redge}). Residuals near 9--10 keV disappeared.
    Dotted lines in each panel show individual model components. See Section \ref{sec:4.1.2}.}
    \label{fig:5-1-efb}
\end{figure*}

\begin{figure*}
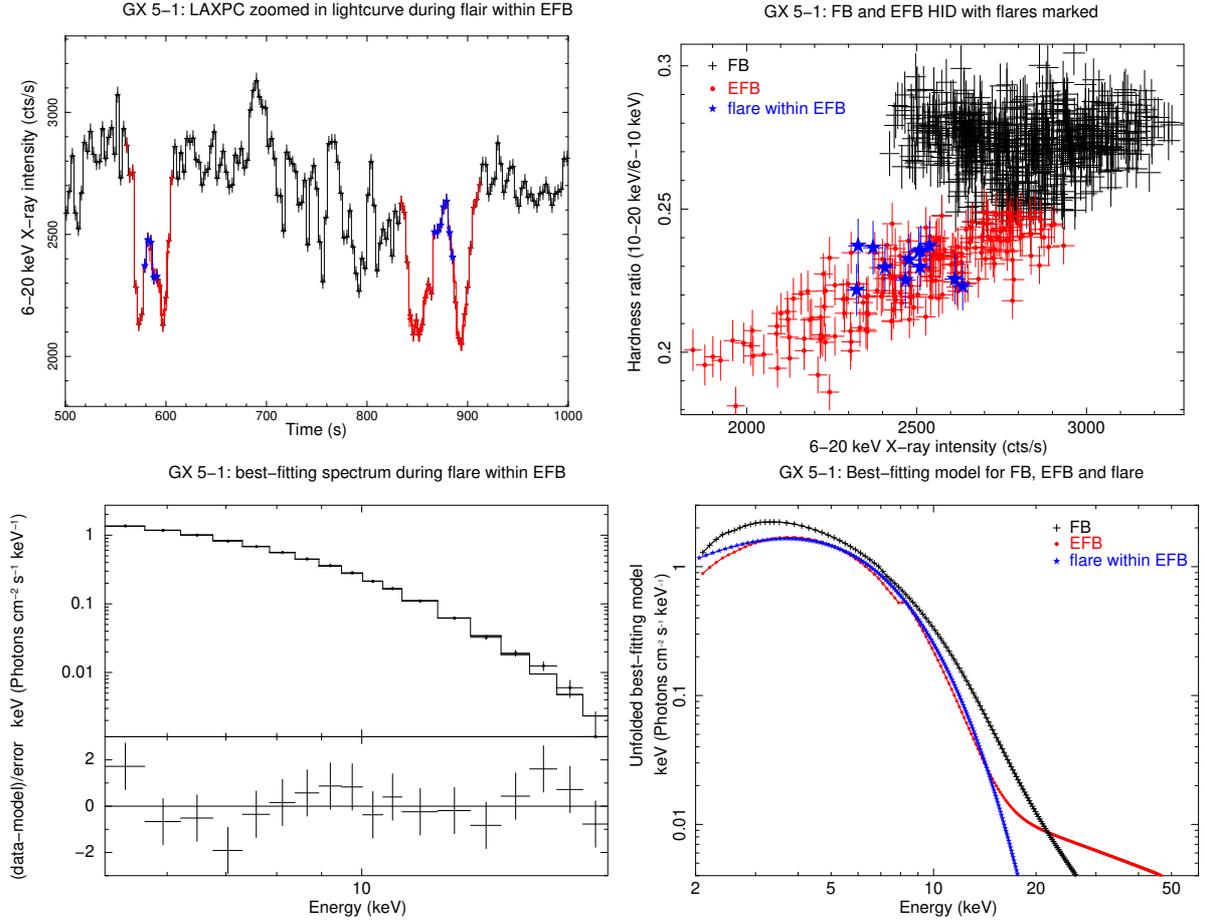

    \centering
    \includegraphics[angle=-90,scale=0.31]{fig5a.eps}
    \includegraphics[angle=-90,scale=0.31]{fig5b.ps}
    \includegraphics[angle=-90,scale=0.31]{fig5c.ps}
    \includegraphics[angle=-90,scale=0.31]{fig5d.ps}
    \caption{Top Left: Zoomed lightcurve of the top left panel in Figure \ref{fb-efb-5-1} is shown in the 400--1000s time range. Two prominent flaring events are observed within the EFB dips and are shown using blue stars. The position of flares is shown in the HID of GX 5-1 during FB and EFB in the top right panel. The best-fitting spectrum, along with the residual of the flaring event (shown by stars in the lightcurve and HID) inside the EFB dips, are shown in the bottom left panel.  The spectrum can be fitted with a single blackbody radiation model. For comparison, best-fitting unfolded models for FB (bottom left panel of Figure \ref{fig:5-1-fb}), EFB (bottom right panel of Figure \ref{fig:5-1-efb}), as well as the flare inside EFB (bottom left panel of the current figure), are shown in the bottom right panel. See Section \ref{sec:4.2}.}
    \label{flare}
\end{figure*}

\begin{table}
\centering
\caption{Best-fitting parameters of spectral analysis of flaring branch in GX~5-1.}
  \begin{tabular}{ccccc}
    \toprule
     & Model & Parameter & ModelA$^a$ & ModelB$^b$  \\
    \midrule
    \label{val:18} & \texttt{Tbabs} & N$_H$(10$^{22}$~cm$^{-2}$) &$3.00^{+0.29}_{-0.27}$& $3.07^{+0.15}_{-0.14}$ \\
    \\
    \label{val:14} & \texttt{Bbodyrad} & k$T_{\rm bb}$ (keV) &$1.36^{+0.01}_{-0.02}$&$1.53^{+0.04}_{-0.03}$ \\
    \\
    \label{val:14} &   & Norm$^c$ &$87^{+6}_{-4}$& $111^{+18}_{-17}$  \\
    \\
        \label{val:14} & \texttt{diskbb} & k$T_{\rm disk}$ (keV) &-& $1.08^{+0.01}_{-0.08}$  \\
    \\
    \label{val:14} &  & norm &-& $464^{+187}_{-141}$  \\
    \\
    \label{val:14} & \texttt{Nthcomp} & Photon Index($\Gamma$) & $5.30^{+0.2}_{-0.4}$& $<$2.36   \\
    \\
    \label{val:14} &    & k$T_{\rm e}$ (keV) & 500(f) & $>$4.53  \\
    \\
    \label{val:14} &    & k$T_{\rm Seed}$(keV) &=k$T_{\rm bb}$& =k$T_{\rm disk}$  \\
    \\
    \label{val:14} &    & Norm (10$^{-2}$) & $180^{+4}_{-3}$ & $1.79^{+0.03}_{-0.15}$ \\
    \\
    \label{val:14} &  & $F_{{\rm Bbodyrad}}$ & $3.72^{+0.11}_{-0.08}$ & $4.38^{+0.23}_{-0.18}$
 \\ 
    & & (10$^{-9}$ ergs/s/cm$^{2}$) & & 
    \\
    \label{val:14} &  & $F_{{\rm Diskbb}}$ & -- & $7.76^{+1.03}_{-0.98}$  \\
        & & (10$^{-9}$ ergs/s/cm$^{2}$) & & 
    \\
        \label{val:14} &  & $F_{{\rm Nthcomp}}$ & $8.51^{+0.17}_{-0.17}$ & $0.03^{+0.01}_{-0.01}$  \\
        & & (10$^{-9}$ ergs/s/cm$^{2}$) & & 
    \\
    \label{val:14} &  & $\chi^2/($dof$)$ &135/123~(1.09)& 118/121~(0.97)  \\
    
    \\
    
    \bottomrule
    \\
  \end{tabular}

   \begin{minipage}{10cm}
{$^a$ \texttt{tbabs}*(\texttt{bbodyrad}\ +\ \texttt{nthcomp})}\\
{$^b$ \texttt{tbabs}*(\texttt{bbodyrad}\ +\ \texttt{diskbb} +\ \texttt{nthcomp})}\\
{$^c$Blackbody normalization is defined as {$R^2/D^2$} where R and D are the source radius \\ and distance in units of km and 10 kpc respectively} \\

\end{minipage}
  \label{tab:5-1-fb}
\end{table}

\begin{figure*}
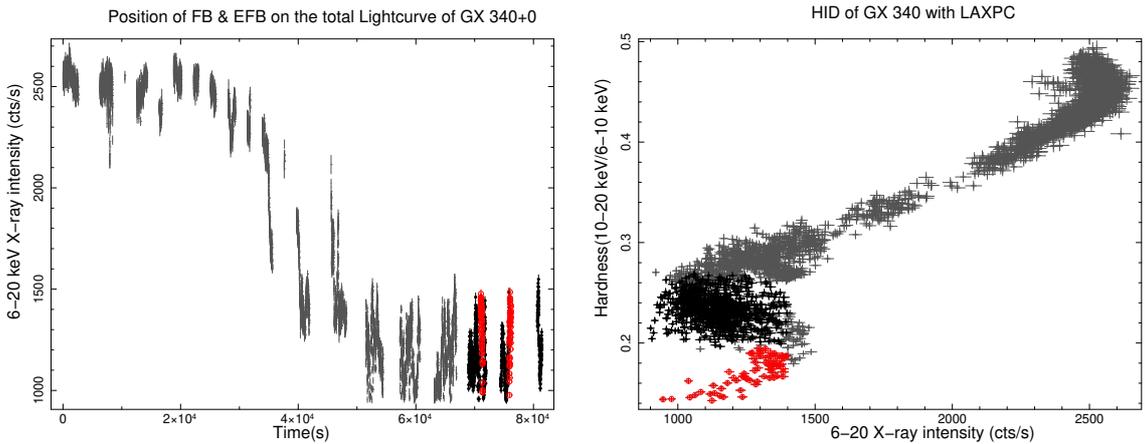

    \centering
    \includegraphics[angle=-90,scale=0.31]{fig6a.eps}
    \includegraphics[angle=-90,scale=0.31]{fig6b.eps}
    \caption{Left: \asat{}/LAXPC light curve of GX~340+0 of FB \& EFB in 6–-20 keV with 5 s bins. Right: the \textit{LAXPC} hardness–intensity diagram (HID) of the GX~340+0 of FB \& EFB  with hard colour computed using the background-subtracted count rates in 10-–20 keV divided by that in 6-–10 keV and intensity has been represented by the total count rate in 6-–20 keV. See Section \ref{sec:4.2}.}
    \label{340-total}
\end{figure*}

\begin{figure*}
    \centering
\includegraphics[angle=-90,scale=0.31]{fig7a.eps}
\includegraphics[angle=-90,scale=0.31]{fig7b.eps}
\includegraphics[angle=-90,scale=0.31]{fig7c.eps}
\includegraphics[angle=-90,scale=0.31]{fig7d.ps}
    \caption{Upper Left: \asat{}/LAXPC light curve of GX~340+0 of FB \& EFB in 6-–20 keV with 5 s bins. Upper Right: the \textit{LAXPC} hardness–intensity diagram (HID) of the GX~340+0 of FB \& EFB  with hard colour computed using the background-subtracted count rates in 10-–20 keV divided by that in 6-–10 keV and intensity has been represented by the total count rate in 6-–20 keV. Bottom left: Variation of the Hardness with respect to time. Bottom right: The portion of 0.3--8 keV \asat{}/SXT lightcurve, which is simultaneous with \textit{LAXPC}, is shown. See Section \ref{sec:4.2}.}
    \label{340-part}
\end{figure*}

\begin{figure*}
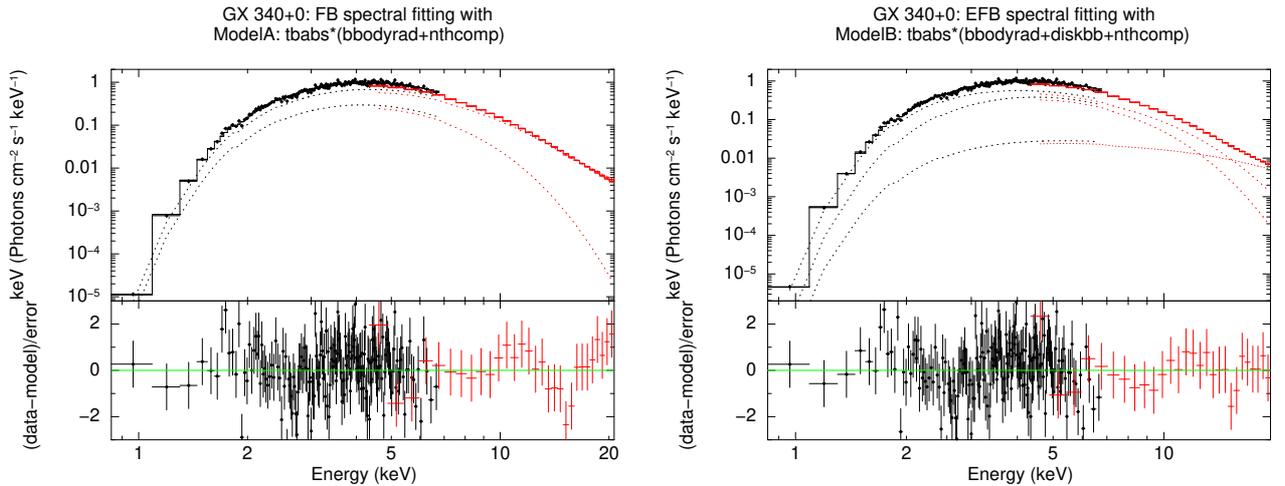

    \centering
\includegraphics[angle=-90,scale=0.31]{fig8a.ps}
\includegraphics[angle=-90,scale=0.31]{fig8b.ps}
\caption{Left: Best-fitting spectra of the flaring branch of GX~340+0. Here, the best-fitting \texttt{XSpec} model is \texttt{TBabs}*(\texttt{nthcomp} + \texttt{bbodyrad}). Right: Best-fitting spectra of the extended flaring branch of GX~340+0. Here, the best-fitting model is \texttt{tbabs}*(\texttt{nthcomp} + \texttt{bbodyrad} + \texttt{diskbb}). The lower panel of both figures represent the residual of the best-fitting model. See Section \ref{sec:4.2.1}.}
    \label{fig:340-fb}
\end{figure*}

\begin{figure*}
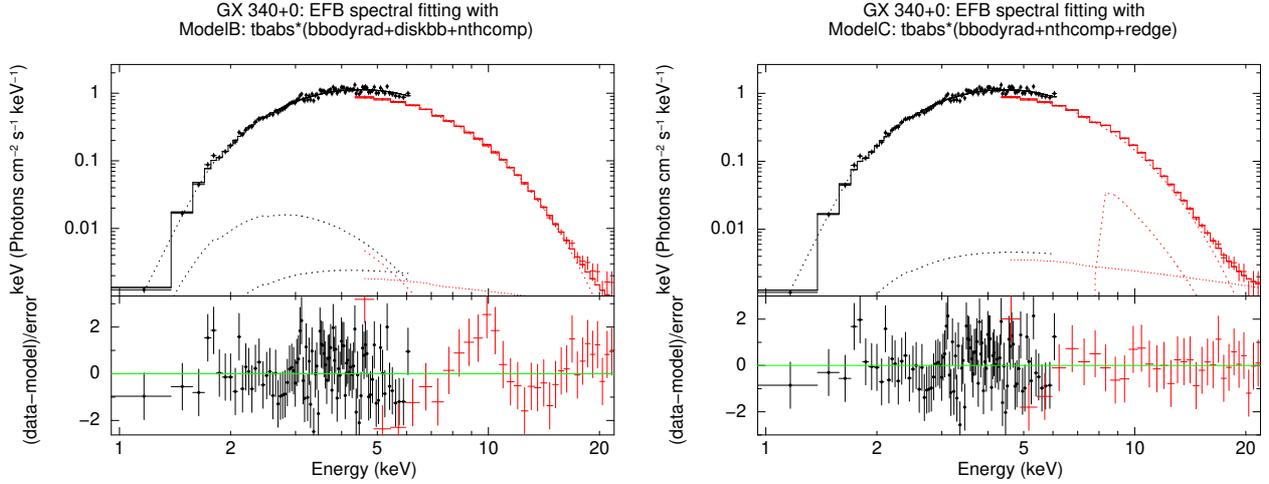

    \centering
\includegraphics[angle=-90,scale=0.31]{fig9a.ps}
\includegraphics[angle=-90,scale=0.31]{fig9b.ps}
\caption{Left: Best-fitting spectra of the extended flaring branch of GX~340+0 without RRC continuum. Here the best-fitting \texttt{XSpec} model is \texttt{tbabs}*(\texttt{nthcomp} + \texttt{bbodyrad}). Right: best-fitting spectra of the extended flaring branch of GX~340+0 with RRC continuum. Here, the best-fitting \texttt{XSpec} model is \texttt{tbabs}*(\texttt{nthcomp} + \texttt{bbodyrad} + \texttt{redge}). The lower panel of both figures represent the residual of the best-fitting model. See Section \ref{sec:4.2.1}.}
    \label{fig:340-efb}
\end{figure*}

\section{Results}
\label{sec:4}
\subsection{GX~5-1}
\label{sec:4.1}
We have considered \asat{}/LAXPC and \asat{}/SXT data from the observation epoch between 03-04 March 2016. In the left panel of Figure \ref{gx51-light-hid}, we have shown the 6--20 keV light curve of GX 5-1 using \textit{LAXPC} observations spanning over $\sim$46 ks, while in the right panel, we have shown the HID of the same span of the lightcurve. NB, FB and EFB are clearly detected in the HID. In both panels, the positions of FB and EFB are shown in black crosses and red circles, respectively. 
As we wish to understand the origin of EFB and its connection to FB, We have shown the zoomed-in portion of the lightcurve and HID, which is part of FB and EFB in the top left and top right panels of Figure \ref{fb-efb-5-1}. The hardness as a function of time is shown in the bottom left panel of Figure \ref{fb-efb-5-1}. A symbol convention similar to Figure \ref{gx51-light-hid} is used. During EFB, the hardness ratio drops significantly by a factor of $\sim$1.5. The bottom right panel of Figure \ref{fb-efb-5-1} shows the simultaneous \textit{SXT} lightcurve in 0.3--7 keV with the bin size of 7s. Due to the lower observational efficiency compared to \textit{LAXPC}, only one dip is observed. Following the colour and symbol convention similar to \textit{LAXPC}, The FB and EFB sections are shown in the \textit{SXT} light curve. 
\subsubsection{FB spectral analysis}
\label{sec:4.1.1}
We have carried out the spectral analysis of FB using joint and simultaneous observations of \textit{SXT} and \textit{LAXPC} in the energy range 0.5--7.0 keV and 3.0--22.0 keV, respectively. Motivated by earlier works, we have fitted joint spectra using a combination of blackbody radiation (\texttt{bbodyrad} in \texttt{XSpec}), thermal comptonised emission from the boundary layer (\texttt{nthcomp} in \texttt{XSpec}) modified by the absorption (\texttt{TBabs}) (ModelA). Such a choice of model components is similar to that used by \citet{church} while analysing {\textsc RXTE} spectra of GX 5-1 during FB and EFB branches. We have obtained the $\chi^2/$dof$=135/123$ (1.09). However, the photon index of the (\texttt{nthcomp} in \texttt{XSpec}) is unusually high ($\sim$5.3), which cannot be explained. A very high photon index may indicate the presence of another missing blackbody-like component. For further justification, we replaced the \texttt{bbodyrad} model with \texttt{diskbb} and kept other model components the same. With the combination of \texttt{TBabs*(diskbb+nthcomp)}, we obtained the $\chi^2/$dof$=137/123$ (1.11) with the disk blackbody normalization of 52$^{+4}_{-6}$. Such a normalization implies an unusually low inner disk radius of 7.7$\pm$0.7 km, which is less than the typical neutron star radius of 10 km, assuming the disk inclination angle of 45$^{\circ}$ and the upper limit of the distance to the source of 9 kpc \citep{christ1997}.
To address the issue further, we replaced the earlier model with another model combining {\texttt bbodyrad} and {\texttt diskbb} models together with {\texttt nthcomp}. Using a combination of \texttt{TBabs*(bbodyrad+diskbb+nthcomp)} (ModelB), we have obtained an acceptable fit with $\chi^2$/dof = 118/121 (0.97). From the best-fitting spectral parameters, the 1$\sigma$ upper limit of the powerlaw index is found to be 2.36, while the disk blackbody normalization is 464$^{+187}_{-141}$. Therefore, \texttt{TBabs*(bbodyrad+diskbb+nthcomp)} (ModelB) is considered the best-fitting model for broadband FB spectral analysis in GX~5-1.
Best-fitting parameters for Models A and B for FB spectral analysis are provided in Table \ref{tab:5-1-fb} along with fluxes and reduced $\chi^2$ values.
\subsubsection{EFB spectral analysis}
\label{sec:4.1.2}
We have adopted a similar strategy for fitting the EFB spectra in GX~5-1. First, we used ModelA for the EFB spectral analysis. With a combination of \texttt{TBabs}*(\texttt{bbodyrad} + \texttt{nthcomp}), we see a strong residual around 8--11 keV. We have replaced the \texttt{bbodyrad} of ModelA with \texttt{diskbb} and also independently used ModelB to fit the spectra. In all three cases, the residual near 8--11 keV persists. The $\ chi^2$/dof values for ModelA and ModelB are 115/86~(1.33) and 112/84~(1.32), respectively. Therefore, Models A and B fail to describe the feature near 8--11 keV. For example, EFB spectra fitted with ModelB are shown in the left panel of fig \ref{fig:5-1-efb} along with the residuals. Hence, the best-fitting model which can adequately describe the FB spectra cannot be used for modelling EFB spectra.

The fitting statistics can be improved significantly by adding a model that describes emission and absorption features near 8--11 keV caused by radiative recombination continuum (\texttt{redge} model in \texttt{XSpec}). When combined with ModelA, the best-fitting model for EFB is \texttt{TBabs}*(\texttt{bbodyrad} + \texttt{nthcomp} + \texttt{redge}) (ModelC). Without (ModelA) and with (ModelC), the RRC model component, the best-fitting $\chi^2$/dof are found to be 115/86 (1.34) and 86/83 (1.04), respectively. An F-test between these two models yields an F statistic value = 9.32 and an F-test probability of 2.21$\times$10$^{-5}$. Therefore, our spectral analysis strongly suggests the presence of RRC during the EFB in GX 5-1. In Table \ref{tab:5-1-efb}, we have provided the best-fitting spectral parameters using models B, A and C, respectively.

By comparing the best-fitting parameters for FB (ModelB in Table \ref{tab:5-1-fb}) and EFB (ModelC in Table \ref{tab:5-1-efb}) spectral fitting, we can see an important change: A significant increase ($\sim$4 times) in the blackbody normalization from 111$^{+18}_{-13}$ to 452$^{+10}_{-13}$ is noted during the transition from FB to EFB. Assuming the distance to the source 9 kpc \citep{christ1997}, the blackbody emitting area is observed to increase from 90.3$^{+1.9}_{-1.7}$~km$^2$ to 361$^{+14}_{-11}$~km$^2$ during the transition from FB to EFB in GX 5-1.

\begin{table*}
\centering
\caption{Best-fitting parameters of spectral analysis of extended flaring branch of GX~5-1. }
  \begin{tabular}{cccccc}
    \toprule
      Model & Parameter & ModelB$^a$ & ModelA$^b$ & ModelC$^c$ &  \\
    \midrule
    \label{val:18}  \texttt{Tbabs} & $N_{\rm H}$(10$^{22}$~cm$^{-2}$) &$2.01^{+0.22}_{-0.41}$&$1.79^{+0.3}_{-0.36}$& $1.96^{+0.23}_{-0.26}$ & \\
    \\
    \label{val:14}  \texttt{Bbodyrad} & k$T_{{\rm bb}}$(keV) &$1.24^{+0.03}_{-0.02}$&$1.26^{+0.02}_{-0.01}$& $1.19^{+0.02}_{-0.01}$ & \\
    \\
    \label{val:14}   & Norm &$384^{+24}_{-25}$&$465^{+19}_{-15}$& $452^{+10}_{-13}$ & \\
    \\
    \label{val:14}  \texttt{diskbb} & k$T_{{\rm disk}}$(keV) & $<$0.72 & -- & -- & \\
    \\
    \label{val:14}   & Norm &$552^{+538}_{-207}$& -- & -- & \\
    \\
    \label{val:14}  \texttt{Nthcomp} & Photon Index($\Gamma$) & $<$2.69 & $2.24^{+0.43}_{-0.44}$& $2.24^{+0.53}_{-0.55}$  & \\
    \\
    \label{val:14}     & k$T_{\rm e}$ (keV) & $>$300 & $>$277 & $>$4.89 & \\
    \\
    \label{val:14}    & k$T_{{\rm Seed}}$ (keV) &=k$T_{{\rm disk}}$ & =k$T_{{\rm bb}}$& =k$T_{{\rm bb}}$ & \\
    \\
    \label{val:14}    & Norm &$0.02^{+0.01}_{-0.01}$ &$0.06^{+0.04}_{-0.03}$& $0.05^{+0.04}_{-0.03}$ & \\
    \\
        \label{val:14}  \texttt{Redge} & edge(keV) &--&--& $8.10^{+0.65}_{-0.62}$  & \\
    \\
    \label{val:14}   & k$T$(keV) &--&--& $1.11^{+0.55}_{-0.53}$  & \\
    \\
    \label{val:14}   & Norm &--&--& $0.08^{+0.01}_{-0.03}$  & \\
    \\
    \label{val:14}   & $F_{{\rm Bbodyrad}}$ & $10^{+2}_{-2}$ & $11^{+2}_{-1}$ & $10^{+1}_{-1}$  & \\
    & (10$^{-9}$ ergs/s/cm$^{2}$) & &
    \\
    \label{val:14}   & $F_{{\rm diskbb}}$ &$1.86^{+0.05}_{-0.04}$&-- & --  & \\
    & (10$^{-9}$ ergs/s/cm$^{2}$) & &
    \\
    \label{val:14}   & $F_{{\rm Nthcomp}}$ &$0.17^{+0.03}_{-0.02}$&$0.81^{+0.05}_{-0.05}$ & $0.78^{+0.04}_{-0.04}$  & \\
    & (10$^{-9}$ ergs/s/cm$^{2}$) & &
    \\
    \label{val:14}   & $F_{{\rm Redge}}$ & -- &--& $0.14^{+0.01}_{-0.01}$  & \\
    & (10$^{-9}$ ergs/s/cm$^{2}$) & &
    \\
    \label{val:14}   & $\chi^2/($dof$)$ &112/84~(1.32)&115/86~(1.34)& 86/83~(1.04) &  \\
    
    \bottomrule
    \\
  \end{tabular}
  \begin{minipage}{10cm}
{$^a$ \texttt{tbabs}*(\texttt{bbodyrad}\ +\ \texttt{diskbb}\ +\ \texttt{nthcomp})}\\
{$^b$ \texttt{tbabs}*(\texttt{bbodyrad}\ +\ \texttt{nthcomp})}\\
{$^c$ \texttt{tbabs}*(\texttt{bbodyrad}\ +\ \texttt{nthcomp}\ +\ \texttt{redge})}\\

\end{minipage}
  \label{tab:5-1-efb}
\end{table*}

\subsubsection{Flaring within EFB}
\label{sec:4.1.3}
A close look at the EFB dips in the top left panel of Figure \ref{fb-efb-5-1} reveals that nearly all dips consist of a sharp flaring activity which lasts from a few seconds to a few tens of seconds. The left panel of Figure \ref{flare} shows the zoomed-in EFB where flares within EFB dips are clearly visible. We have created a GTI file with all flare intervals shown by stars in the top left panel of Figure \ref{flare}. Flare interval is defined in the time range of $\pm$10s around flare peak time. These criteria ensure the exclusive selection of active flaring regions with good data statistics. The position of flares is shown in the HID of FB and EFB in the top right panel of Figure \ref{flare}.
Using the selected GTI, we have extracted \textit{LAXPC20} source and background spectra in 3--15 keV and have used them for further analyses. We found that background-subtracted spectrum can be fitted by a single blackbody radiation model \texttt{bbodyrad} modified by absorption with the best-fitting $\chi^2$/dof = 15/16. The absorption is kept fixed at 1.96 $\times 10^{22} cm^{-2}$, which is the best-fitting value obtained from joint \textit{SXT} and \textit{LAXPC} spectral modelling provided in Table \ref{tab:5-1-efb}. The best-fitting blackbody temperature and normalization are found to be 1.31$^{+0.03}_{-0.02}$ keV and 538$^{+13}_{-12}$ respectively. Interestingly, the blackbody emitting area is larger by a factor of $\sim$2.5 compared to that observed during FB. Therefore, such a rapid expansion of the blackbody emitting region occurs in a timescale of a few tens of seconds, which is interesting. A similar rapid expansion of blackbody radius has been observed from the time-resolved spectral analysis of type-I X-ray bursts from neutron star surfaces of different NSXBs \citep{kuulkers2003,galloway2008,beri2019}. Short bursts, of the order of 30-50 sec, are explained in terms of either a pause in nuclear chain reaction \citep{fisker2004} or by convective transportation of leftover material to the ignition depth from the previous burst \citep{keek2017}. However, we may observe that the profile of type-I burst and the EFB flare are different: type-I bursts have a sharp rise ($\leq$ 2sec) and exponential fall (10 sec or more) \citep{galloway2008} while EFB flares are more symmetric with the rise and fall time around 20-30 sec. Therefore, it is unclear whether type-I X-ray bursts and observed EFB flares have the same origin, which is subject to further investigation and out of the scope of current work. We have extracted 0.1-1000 Hz power density spectra during burst intervals in the energy range of 3--80.0 keV. No high-frequency burst oscillations have been detected with the 0.1-1000.0 Hz integrated fractional rms upper limit of 4.47\%.
For comparison, the best-fitting model parameter values for the FB, EFB and flares are provided in Table \ref{tab:comparison}, and best-fitting unfolded models are shown in the bottom right panel of Figure \ref{flare}. As the table suggests, the FB has a higher total flux than the EFB, while the flare within the EFB has the highest Bolometric flux. Moreover, a comptonization tail is evident during EFB, which is absent from FB. Therefore, our spectral modelling of FB, EFB and flare suggests that EFB is not an extension of FB. 
For FB, EFB and flare, We have calculated $L/L_{Edd}$ using best-fitting flux in 0.01--100 keV (calculated using {\textsc cflux} model in \texttt{XSpec}) and assuming the neutron star mass of 1.7 \(\textup{M}_\odot\) and a distance of 9 kpc \citep{christ1997}. The $L/L_{Edd}$ values are found to be 80$^{+4}_{-5}$\%, 74$^{+3}_{-3}$\% and 96$^{+6}_{-5}$\% during the FB, EFB and flare respectively. 
Assuming an upper and lower limit of the neutron star mass of 2.0\(\textup{M}_\odot\) and 1.5\(\textup{M}_\odot\) respectively, we have determined $L/L_{Edd}$ within the range of 66--98\%, 58--83\% and 78--123\% for the EFB and flare respectively.

\begin{table*}
\centering
\caption{Comparison of best-fitting parameters from spectral analysis of EFB and EFB flare in GX~5-1. }
  \begin{tabular}{ccccc}
    \toprule
   Model & Parameter  & EFB & EFB flare \\
     &  & (\textit{SXT}+\textit{LAXPC}) & \textit{LAXPC} \\
    \midrule
    \label{val:18}  \texttt{Tbabs} & $N_{\rm H}$(10$^{22}$~cm$^{-2}$) & $1.96^{+0.23}_{-0.26}$ & 1.96(f)\\
    \\
    \label{val:14}  \texttt{Bbodyrad} & k$T_{{\rm bb}}$(keV)  & $1.19^{+0.02}_{-0.01}$ & $1.31^{+0.03}_{-0.02}$ \\
    \\
    \label{val:14}   & Norm & $452^{+10}_{-13}$ & $538^{+13}_{-12}$\\
    \\
    \label{val:14}  \texttt{Nthcomp} & Photon Index($\Gamma$) & $2.24^{+0.53}_{-0.55}$  & -- \\
    \\
    \label{val:14}   & k$T_{\rm e}$(keV)$^d$  & $>4.89$ & -- \\
    \\
    \label{val:14}     &=k$T_{{\rm bb}}$& =kT$_{{bb}}$ & -- \\
    \\
    \label{val:14}    & Norm & $0.05^{+0.04}_{-0.03}$ & --\\
    \\
        \label{val:14}  \texttt{Redge} & edge(keV) & $8.10^{+0.65}_{-0.62}$  & -- \\
    \\
    \label{val:14}  & k$T$(keV) & $1.11^{+0.55}_{-0.53} $ & -- \\
    \\
    \label{val:14}  & norm & $0.08^{+0.01}_{-0.03}$  & -- \\
    \\
    \label{val:14}   & $F_{{\rm Bbodyrad}}$ &  $10^{+1}_{-1}$  & $14^{+1}_{-1}$ \\
    & (10$^{-9}$ ergs/s/cm$^{2}$) & & &
    \\
    \label{val:14}   & $F_{{\rm Nthcomp}}$ &  $0.78^{+0.04}_{-0.04}$  & -- \\
    & (10$^{-9}$ ergs/s/cm$^{2}$) & & &
    \\
    \label{val:14}  & $F_{{\rm Redge}}$  & $0.14^{+0.01}_{-0.01} $ & -- \\
    & (10$^{-9}$ ergs/s/cm$^{2}$) & & &
    \\
    \\
    \label{val:14}   & $\chi^2/($dof$)$ & 86/83~(1.04) & 14/13~(1.07) \\
    \\
    \label{val:14}   & $L/L_{\mathrm{Edd}}^d$ (\%) & $62.9^{+2.2}_{-1.7}$ &$ 79.3^{+3.1}_{-2.9}$ \\    
    \bottomrule
    \\
  \end{tabular}
  \begin{minipage}{10cm}
{$^d$ $L$ is the Bolometric luminosity calculated in the energy range 0.01--100.0 keV using best-fitting model parameters, and $L_{\mathrm{Edd}}$ is the Eddington luminosity.}

\end{minipage}
  \label{tab:comparison}
\end{table*}

\subsection{GX~340+0}
\label{sec:4.2}
The spectral properties of HB and NB were studied in detail by \citet{yash2023} using \asat{}/LAXPC and \asat{}/SXT, while the timing analysis was performed by \citet{pahari2024astrosat}. Here, we have used the \asat{} data from the epoch, the same as \citet{pahari2024astrosat}, since it shows the presence of EFB. The left panel of Figure \ref{340-total} shows the 6--20 keV light curve of the source using \textit{LAXPC} observations, while the right panel shows the HID. In both panels, the positions of FB and EFB are shown in black pluses and red circles. The zoomed-in portion of the lightcurve and HID that belongs to FB and EFB are shown in the top left and top right panels of Figure \ref{340-part}. The hardness as a function of time is shown in the bottom left panel of Figure \ref{340-part}. During EFB, the hardness drops significantly. The bottom right panel of Figure \ref{340-part} shows the simultaneous \textit{SXT} lightcurve in 0.3--8 keV with a bin size of 7s. Following the colour and symbol convention similar to \textit{LAXPC} analysis, The FB and EFB are shown in the \textit{SXT} light curve.

\subsubsection{FB and EFB Spectral analysis}
\label{sec:4.2.1}
Following the strategy similar to GX 5-1, we have carried out the spectral analysis of FB and EFB in GX 340+0 separately using joint observation of \textit{SXT} and \textit{LAXPC} in the energy range (0.5--22 keV). The left panel of Figure \ref{fig:340-fb} shows the best-fitting spectra of FB along with the residual when the spectrum was fitted with ModelA: a combination of blackbody radiation (\texttt{bbodyrad} in \texttt{XSpec}), thermal comptonised emission from the boundary layer (\texttt{nthcomp} in \texttt{XSpec}) modified by the absorption (\texttt{TBabs}). The fit returned an acceptable $\chi^2$/dof = 217/170 (1.28). However, the fit returned an unusually high photon index of the thermal comptonization model: 4.55$^{+0.20}_{-0.18}$. The issue is similar to that observed during the FB spectral analysis of GX 5-1 and has not been resolved even with replacing \texttt{bbodyrad} with \texttt{diskbb}.

Therefore, we use ModelB: \texttt{TBabs*(diskbb+bbodyrad+nthcomp)} to fit the \textit{SXT+LAXPC} joint spectra. There is a marginal improvement in model fitting with $\chi^2$/dof = 199/168(1.18); moreover, the 1$\sigma$ upper limit of the photon index is found to be 2.19. The electron temperature and disk blackbody model parameters are also constrained and within an acceptable range. Therefore, similar to GX 5-1, ModelB is found to be the best-fitting model for the broadband spectral analysis of FB in GX 340+0. The best-fitting parameters of ModelA and ModelB are provided in Table \ref{tab:340-flaring}.

We may note that using a combination of blackbody radiation {\texttt bbodyrad} and thermal comptonization {\texttt thcomp}, \citet{su2024} carried out the `Z' track-resolved spectral analysis of GX 340+0 using \textit{SXT} and \textit{LAXPC} observations of GX 340+0. Spectral parameters corresponding to the FB branch are similar to what we obtained in the present work.

A strategy similar to EFB spectral modelling of GX~5-1 has been adopted to fit the EFB spectra in GX 340+0. First, we have used ModelA and ModelB to fit EFB spectra from GX 340+0. Best-fitting spectral parameters are shown in Table \ref{tab:340-efb}. Fitting with ModelA and ModelB returned similar goodness of fit with $\chi^2$/dof = 153/122 (1.25) and 149/120 (1.24) respectively. For either case,  strong residuals have been observed around 8--11 keV in the left panel of Figure \ref{fig:340-efb}. A significant improvement in the fit statistics has been observed when the RRC model component \texttt{redge} is added with ModelA and the spectra are fitted with ModelC: \texttt{TBabs*(bbodyrad+nthcomp+redge)}. Due to the addition of \texttt{redge} to ModelA, $\chi^2$/dof changes from 153/122 (1.25) to 123/119 (1.03). The RRC energy and electron temperature are kept free to vary. An F-test between these two models yields an F statistic value = 9.68 and an F-test probability of 9.18 $\times$ 10$^{-6}$. The best-fitting RRC energy and temperature are found to be 7.91$^{+0.25}_{-0.22}$ keV and 1.25$^{+0.45}_{-0.43}$ keV respectively. Therefore, energy spectral analysis supports the presence of RRC during the EFB in GX 340+0. Table \ref{tab:340-efb} provides the best-fitting spectral parameters with the RRC model component (ModelC). Two important changes have been noted when best-fitting FB and EFB parameters are compared from table \ref{tab:340-flaring} and table \ref{tab:340-efb}: A significant increase in the blackbody normalisation (indicating blackbody emitting area) from 173$^{+36}_{-57}$ to 411$^{+12}_{-11}$ during the transition from FB to EFB.

\begin{table}
\centering
\caption{Best-fitting parameters of spectral analysis of flaring branch of GX~340+0. }
\begin{tabular}{cccc}
\toprule
Model & Parameter & Model$A^a$  & ModelB$^b$\\
\midrule
\label{val:18}  ${\texttt Tbabs}$ & $N_{\rm H}$(10$^{22}$~cm$^{-2}$)  & $5.27^{+0.16}_{-0.17}$ &$5.09^{+0.23}_{-0.21}$ \\
\\
\label{val:14}  ${\texttt Bbodyrad}$ & k$T_{{\rm bb}}$(keV) &  $1.28^{+0.03}_{-0.05}$ & $1.29^{+0.02}_{-0.03}$ \\
\\
\label{val:14}    & Norm$^c$  & $65^{+3}_{-5}$ &$173^{+36}_{-57}$\\
\\
\label{val:14}  ${\texttt diskbb}$ & k$T_{\rm disk}$ (keV) &--& $1.22^{+0.03}_{-0.03}$   \\
\\
\label{val:14}   & Norm &--& $86^{+16}_{-10}$   \\
\\
\label{val:14}  ${\texttt Nthcomp}$ &  Photon Index($\Gamma$) & $4.55^{+0.20}_{-0.18}$  & $<$2.19 \\
\\
\label{val:14}     & k$T_{\rm e}$(keV) & 500(f) & $3.98^{+0.9}_{-0.8}$ \\
\\
\label{val:14}     & k$T_{\rm Seed}$(keV) & =k$T_{{\rm bb}}$ &  =k$T_{\rm disk}$ \\
\\
\label{val:14}     & Norm & $1.05^{+0.18}_{-0.15}$ & $0.19^{+0.04}_{-0.03}$ \\
\\

\label{val:14}   & $F_{\rm Bbodyrad}$ & $8.91^{+0.36}_{-0.36}$ & $9.98^{+0.66}_{-0.71}$   \\
& & (10$^{-9}$ ergs/s/cm$^{2}$) &  
\\
\label{val:14}   & $F_{\rm diskbb}$ & -- & $2.56^{+0.11}_{-0.09}$   \\
& & (10$^{-9}$ ergs/s/cm$^{2}$) &  
\\
\label{val:14}   & $F_{\rm Nthcomp}$ & $3.39^{+0.14}_{-0.10}$ & $0.06^{+0.01}_{-0.01}$  \\
& & (10$^{-9}$ ergs/s/cm$^{2}$) &  
\\
\label{val:14}   & $\chi^2/(dof)$ & 217/170~(1.28) & 199/168~(1.18)  \\
\bottomrule    
\end{tabular}
\begin{minipage}{10cm}
{$^a$ \texttt{tbabs}*(\texttt{bbodyrad}\ +\ \texttt{nthcomp})}\\
{$^b$ \texttt{tbabs}*(\texttt{bbodyrad}\ +\ \texttt{diskbb} +\ \texttt{nthcomp})}\\
{$^c$Normalization is defined as {$R^2/D^2$}.Where R and D are the source radius and \\
distance in units of  km and 10 kpc respectively} 
\end{minipage}
  \label{tab:340-flaring}
\end{table}

\section{Significance testing of the RRC component}
\label{sec:5}
\begin{figure*}
    \centering
    \includegraphics[width=8.5cm, height=5.5cm]{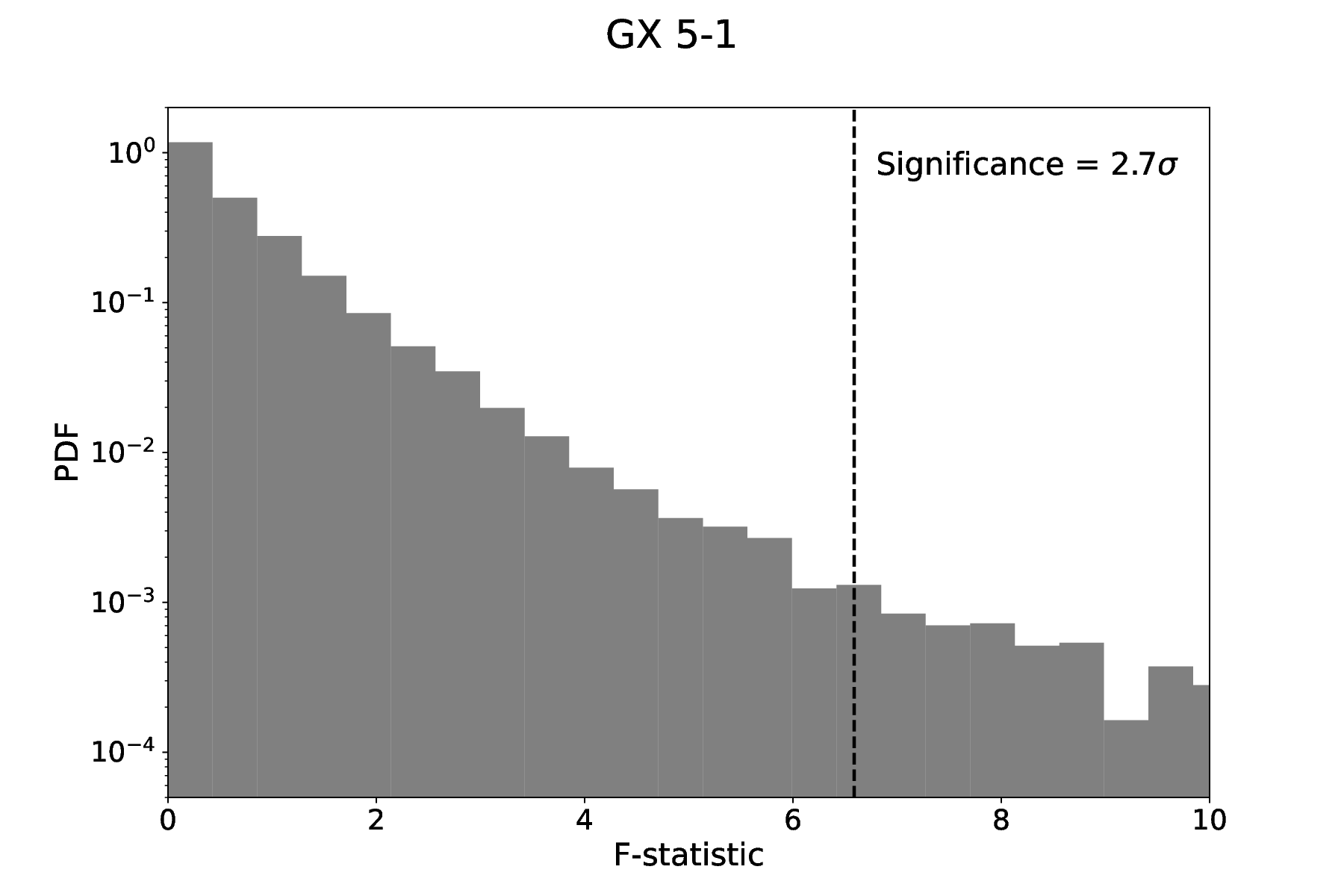}
    \includegraphics[width=8.5cm, height=5.5cm]{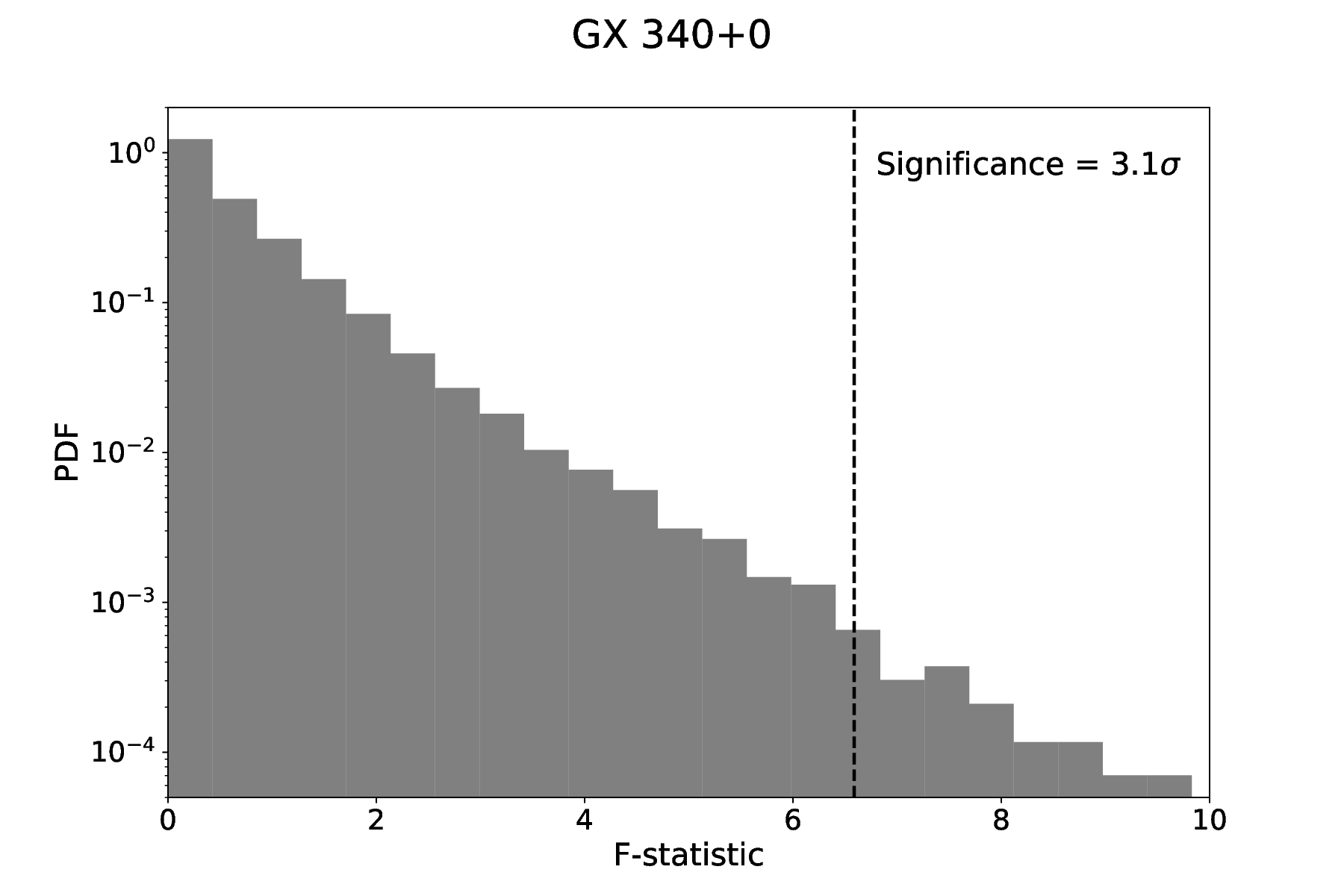}
    \caption{Probability density functions (pdf) on both panels depict the simulated alternative distribution of F-test statistics (i.e., F-test statistics distribution corresponds to ModelC) computed in GX 5-1 (left panel) and GX 340+0 (right panel). The sigma values of the significance of ModelC with respect to ModelA are indicated by the vertical dotted line, which corresponds to the F-statistic cut-off of our simulation. See Section \ref{sec:5}.}
    \label{ftest}
\end{figure*}

To assess the statistical significance of the presence of the RRC component during EFB in GX~5-1 and GX~340+0, we have used a statistical test based on Bayesian posterior predictive probability values following the prescription of \citet{prota02}. For this purpose, we have used ModelC, i.e., model with \texttt{redge} component and model without \texttt{redge} component, i.e., ModelA. Choices of ModelC and ModelA are appropriate because ModelC = ModelA + \texttt{redge}. For significance testing purposes, we have considered ModelA rather than ModelB since ModelB is the best-fitting model for FB spectral analysis only and not for EFB. Secondly, ModelB + \texttt{redge} (ModelD) significantly over-fit the EFB spectra, e.g., in the case of GX 340+0, best-fitting with ModelD yields a $\chi^2/dof$ = 96/117 (0.82).
\begin{itemize}
\item For each source, 100,000 fake spectra (N) have been generated for each of the null models without the \texttt{redge} component (ModelA: \texttt{tbabs}*(\texttt{bbodyrad} + \texttt{nthcomp})) and the alternative model that includes the \texttt{redge} component (ModelC: \texttt{tbabs}*(\texttt{bbodyrad} + \texttt{nthcomp} + \texttt{redge})).
100,000 fake spectra (using \texttt{fakeit} in \texttt{XSpec}) are generated using the best-fitting parameter values for the best-fitting ModelA and ModelC, considering their errors when applying counting statistics. Suitable background spectra and response files are used.

\item For each of the two models,  the F-test was carried out for all the 100,000 spectra, and the corresponding F-statistic was obtained. 

\item The posterior predictive $p$ value has been obtained by setting the boolean values in correspondence to the F-statistic value where the F-statistic values greater than the F-statistic computed from the original data set was assigned a value of 1 and the rest were 0. The mean of this boolean array has been calculated to find the posterior predictive p-value. This is done for both models.
The p-value is then converted to $\sigma$ significance using the following method suggested by \cite{frederic06}

\begin{equation}
\label{eq:significance}
\sigma=\Phi^{-1}(1-p/2)
\end{equation}

where $p$ denotes the p-value and $\Phi^{-1}$ is the inverse of the cumulative distribution function.

\item If the posterior predictive p-value is very small (near 0), it indicates that the ModelA should be rejected in favour of the ModelC. Conversely, if the p-value is sufficiently high (close to 1), it suggests that the ModelC should be rejected and the ModelA can be valid.
\end{itemize}
In the case of GX 5-1, for ModelA, a p-value of 0.0039 is obtained. Similarly, for GX~340+0, for the ModelA, a p-value of 0.00088 clearly shows that the model without \texttt{redge}, i.e., ModelA, can be rejected for the EFB spectral analysis. Following equation \ref{eq:significance}, such p-values correspond to 2.7$\sigma$ and 3.1$\sigma$, respectively.
Hence, statistical analysis shows that the modelling requires an additional \texttt{redge} component, and the residual around 8-9 keV is not due to a statistical fluctuation.
Both panels of Figure \ref{ftest} show the probability density function for the F-statistic distribution of ModelC, i.e., the model with the \texttt{redge} component in GX 5-1 (left panel) and GX 340+0 (right panel) and the corresponding significances.   

\begin{table*}
\centering
\caption{Best-fitting parameters of spectral analysis of extended flaring branch of GX~340+0}
  \begin{tabular}{ccccccc}
    \toprule
     & Model & Parameter & ModelB$^a$ & ModelA$^b$  & ModelC$^c$ &\\
    \midrule
    \label{val:18} & \texttt{Tbabs} & $N_{\rm H}$(10$^{22}$~cm$^{-2}$)  & $3.94^{+0.36}_{-0.28}$ &$4.00^{+0.16}_{-0.17}$ &$4.16^{+0.16}_{-0.17}$& \\
    \\
    \label{val:14} & \texttt{Bbodyrad} & k$T_{{\rm bb}}$(keV) &  $1.32^{+0.02}_{-0.03}$ & $1.30^{+0.01}_{-0.04}$ & $1.25^{+0.01}_{-0.04}$& \\
    \\
    \label{val:14} &   & Norm$^c$ & $426^{+18}_{-13}$ &$408^{+10}_{-15}$ &$410^{+12}_{-11}$& \\
    \\
    \label{val:14} & \texttt{diskbb} & k$T_{{\rm disk}}$(keV) &  $<$0.35 & -- & -- & \\
    \\
    \label{val:14} &   & Norm$^c$ & $<$ 33 & -- &-- & \\
    \\
    \label{val:14} & \texttt{Nthcomp} & Photon Index($\Gamma$) & $<$1.87 & $2.10^{+0.20}_{-0.18}$  & $2.10^{+0.20}_{-0.18}$ &\\
    \\
    \label{val:14} &    & k$T_{\rm e}$(keV) & $>$215 & 500(f) & 500(f) & \\
    \\
    \label{val:14} &    & k$T_{{\rm seed}}$(keV) & =k$T_{{\rm disk}}$ & =k$T_{{\rm bb}}$ & =k$T_{{\rm bb}}$ &\\
    \\
    \label{val:14} &    & Norm & $0.0002^{+0.0001}_{-0.0001} $& $0.006^{+0.01}_{-0.01} $& $0.011^{+0.18}_{-0.15}$ & \\
    \\
    
    \\
    \label{val:14} & \texttt{Redge} & edge(keV) & -- & -- & $7.91^{+0.25}_{-0.22}  $& \\
    \\
    \label{val:14} &  & k$T$(keV) & -- & -- & $1.25^{+0.45}_{-0.43} $ & \\
    \\
    \label{val:14} &  & norm & -- & -- & $0.16^{+0.05}_{-0.04}$  & \\
    \\
    \label{val:14} &  & $F_{{\rm Bbodyrad}}$ $12.6^{+1}_{-1}$  & $12.2^{+1}_{-1}$ & $12^{+1}_{-1}$  & \\
    & & (10$^{-9}$ ergs/s/cm$^{2}$) & & 
    \\
    \label{val:14} &  & $F_{{\rm disk}}$ & $1.12^{+0.11}_{-0.18}$ & -- & --  & \\
    & & (10$^{-9}$ ergs/s/cm$^{2}$) & & 
    \\
    \\
    \label{val:14} &  & $F_{{\rm Redge}}$ & -- & -- & $0.12^{+0.01}_{-0.01}$  & \\
    & & (10$^{-9}$ ergs/s/cm$^{2}$) & & 
    \\
    \label{val:14} &  & $F_{{\rm Nthcomp}}$ & $0.13^{+0.04}_{-0.05}$ & $1.02^{+0.03}_{-0.04}$ & $0.93^{+0.02}_{-0.02}$  & \\
    & & (10$^{-10}$ ergs/s/cm$^{2}$) & & 
    \\
    \label{val:14} &  & $\chi^2/($dof$)$ & 149/120~(1.24) & 153/122~(1.25) &  123/119~(1.03) &  \\
    \bottomrule
    \\
  \end{tabular}
  \begin{minipage}{10cm}
{$^a$ \texttt{tbabs}*(\texttt{bbodyrad}\ +\ \texttt{diskbb}\ +\ \texttt{nthcomp})}\\
{$^b$ \texttt{tbabs}*(\texttt{bbodyrad}\ +\ \texttt{nthcomp})}\\
{$^c$ \texttt{tbabs}*(\texttt{bbodyrad}\ +\ \texttt{nthcomp} + \text tt{redge})}\\
{$^c$Normalization is defined as {$R^2/D^2$} where R and D are the source radius and \\
distance in units of  km and 10 kpc respectively} \\

\end{minipage}
  \label{tab:340-efb}
\end{table*}

\section{Discussion and Conclusion}
\label{sec:6}
In this work, we have revisited the origin of the extended flaring branch or EFB, an additional branch observed in the HID of a few `Z' type NSXBs at the end of their flaring branch (FB). During the monitoring campaign with \asat{}, EFB has been observed from two `Z' type NSXBs: GX 340+0 and GX 5-1. Here, we present the 0.5--22 keV broadband spectral analysis of FB and EFB using joint observations from \textit{SXT} and \textit{LAXPC} onboard \asat{}. Broadband spectra during FB can be well described by combining blackbody radiation from the NS surface and thermal comptonization from the boundary layer. However, during EFB, a strong residual is observed between 8--11 keV, which requires an additional spectral component to describe satisfactorily: radiative recombination. The radiative recombination continuum consists of continuum emission caused by the expansion of boundary layers or unstable nuclear burning on the NS surface, along with strong absorption edges caused by adiabatic cooling. 
In a thermally hot plasma with low optical density and temperature ($\sim$0.1 keV) and which is not in ionization equilibrium, the prominence or absence of RRC features depends on the plasma\'s conditions. A plasma undergoing ionization (e.g., due to recent shock heating) will exhibit faint RRC emission because few highly ionized ions are available for recombination. Conversely, a hot plasma ($\sim$2--3 keV) undergoing recombination (such as when rapidly cooled by expansion) will display robust RRC characteristics as most ions are in the process of recombining. Our results from the spectral analysis support the latter case. 

 Our lightcurve analysis from both sources shows that within FB, EFB has originated, and it is similar to the absorption features/dips occurring within a timescale of a few seconds to tens of seconds (see Figures \ref{fb-efb-5-1}, \ref{flare} and \ref{340-part}). Two possible origins of such dip-like features in the light curve could be (1) clumpy outflowing disk wind from the system causing ionized absorption-induced variability; similar absorption dips have been observed from XV 1323-619 \citep{balu1999}, H 1743-322 \citep{miller2006}, Cyg X-2 \citep{bal2011} or (2) sudden radiation pressure-driven expansion and rapid cooling of an emitting region close to the NS surface \citep{pike2021} or rapid collapse of corona/boundary layer due to cooling and bloated inner-disk due to enhanced accretion rate \citep{bal2023}.
Individual spectral analysis of FB and EFB suggests that there are no significant differences in the broadband absorption properties of FB and EFB. The moderate absorption column density of $\sim$4-5 $\times$ 10$^{22}$ cm$^{-2}$ is observed during both branches in GX 340+0 while the same in the range of $\sim$1.7--3 $\times$ 10$^{22}$ cm$^{-2}$ is observed in both branches of GX 5-1. 
Such an analysis was not possible with earlier studies due to the absence of spectral information below 3 keV. Therefore, the hypothesis that sudden and significant changes in the local absorption cause the EFB dip can be ruled out. Moreover, comparing FB and EFB spectral fitting of Cyg X-2 observations using \textit{XMM-Newton} grating spectra \citep{bal12} shows that the occurrence of EFB neither explicitly depends on inclination-dependent absorption nor on the monotonous increase in mass accretion rate across all branches. Therefore, it is likely that the spectral components present during FB will be different than the spectral components present in the EFB.  

Hence, our second hypothesis is that the presence of an additional physical process causing radiation-driven rapid expansion (since the L/L$_{Edd}$ can be 100\% or higher) and subsequent cooling may play an important role in explaining the occurrence of EFB. We found that the model describing the radiation recombination continuum (\texttt{redge} in \texttt{XSpec}) best describes the process causing residual in 8--11 keV in both sources.  
From the best-fitting spectral parameters shown in Tables \ref{tab:5-1-efb} and \ref{tab:340-efb}, edge energies observed from both sources during EFB are 7.91$^{+0.21}_{-0.15}$ keV and 8.11$^{+0.17}_{-0.16}$ keV, respectively. Therefore, such line energies are consistent with the absorption edges due to Fe XVII, Fe XVIII, Fe XIX and Fe XX. 
Using our spectral analysis, we have computed the blackbody radius ($R_{bb}$), which is related to the \texttt{bbodyrad} model normalisation ($N_{bbodyrad}$) by: 
$N_{bbodyrad}$ = $R_{bb}^2$/$D_{10}^2$, where $R_{bb}$ is the blackbody radius in km, and $D_{10}$ is the distance to the source in the unit of 10 kpc. 
Assuming the distance to GX 5-1 as 9 kpc \citep{christ1997} and using best-fit normalization from Table \ref{tab:5-1-fb} and \ref{tab:5-1-efb}, we have found that the blackbody radius increases from 9.5$^{+0.8}_{-0.7}$ km to 19$^{+2}_{-1}$~km while transiting from FB to EFB in GX 5-1. Additionally, we have observed short timescale flares within EFB of GX 5-1. 
Spectral analysis shows that the flare spectrum within the EFB can be fit with only blackbody emission (with the blackbody radius of 21$^{+2}_{-1}$ km). On the other hand, EFB spectral fitting needs two more components than EFB flares. This shows that EFB-flare may be a separate Z-source spectral state, which was not reported earlier. During the flare and EFB, the radius of the blackbody emission increases nearly by a factor of 2 when compared to that of FB.

A similar change in the radius of the blackbody-emitting component is observed in GX 340+0. Assuming a distance of 11$\pm$3 kpc \citep{fender2000} and using best-fit blackbody normalizations from ModelB in Table \ref{tab:340-flaring} and ModelC in Table \ref{tab:340-efb}, we found that the blackbody radius increases from 14$^{+5}_{-6}$ km to 23$^{+6}_{-7}$ km while transiting from FB to EFB.

There could be one of the following two origins of the blackbody-emitting flare: (I) Thermonuclear burning on the neutron star, where the source of energy is nuclear fusion \citep{bildsten2000,bhattacharyya2022} (II) The emission from the boundary layer (accreted material between the accretion disk and the neutron star), where the source of energy is the gravitational potential energy \citep{gilfanov2014}. The first scenario is clearly not possible because, given that there is no burst observed in the FB and most of EFB, a nuclear burning would be a stable nuclear burning. In that case, there is no reason that the blackbody normalization (proportional to the blackbody emission area) would significantly increase during EFB when the luminosity decreases. Therefore, an increase in the boundary layer volume(and hence the emission area) may explain the observed EFB flare. During the flare,
both the boundary layer volume and temperature increase compared to EFB (which explains the observed higher luminosity). This perhaps affects the Comptonization and RRC model components, making them undetectable in the spectrum. More high-resolution X-ray grating spectroscopic observations during EFB can provide more details of the geometry associated with the occurrence of the radiative recombination phenomenon and associated EFB. However, such an analysis is currently out of the scope of the present work.

\section*{Acknowledgements}
The present work makes use of data from the \asat{} mission of the Indian Space Research Organisation (ISRO), archived at the Indian Space Science Data Centre (ISSDC). This work has been performed utilizing the calibration databases and auxiliary analysis tools developed, maintained, and distributed by \asat{}/LAXPC teams with members from various institutions in India and abroad. This work has used data from the Soft X-ray Telescope (\textit{SXT}), developed at TIFR, Mumbai, and the \textit{SXT} POC at TIFR is thanked for verifying and releasing the data via the ISSDC data archive and providing the necessary software tools.

\section*{Data Availability}
The AstroSat data can be downloaded from  \url{https://astrobrowse.issdc.gov.in/astro archive} using the observation IDs mentioned in Table~\ref{tab:1}.
%%%%%%%%%%%%%%%%%%%%%%%%%%%%%%%%%%%%%%%%%%%%%%%%%%

%%%%%%%%%%%%%%%%%%%% REFERENCES %%%%%%%%%%%%%%%%%%

\label{lastpage}

\end{document}